\documentstyle[12pt]{article}

\newcommand{\nc}{\newcommand}
\nc{\beq}{\begin{equation}}
\nc{\eeq}{\end{equation}}
\nc{\beqa}{\begin{eqnarray}}
\nc{\eeqa}{\end{eqnarray}}

\input epsf
\newwrite\ffile\global\newcount\figno \global\figno=1

\def\writedef#1{}
\def\figin{\epsfcheck\figin}\def\figins{\epsfcheck\figins}
\def\epsfcheck{\ifx\epsfbox\UnDeFiNeD
\message{(NO epsf.tex, FIGURES WILL BE IGNORED)}
\gdef\figin##1{\vskip2in}\gdef\figins##1{\hskip.5in}
\else\message{(FIGURES WILL BE INCLUDED)}%
\gdef\figin##1{##1}\gdef\figins##1{##1}\fi}
\def\figinsert{}
\def\ifig#1#2#3{\xdef#1{fig.~\the\figno}
\writedef{#1\leftbracket fig.\noexpand~\the\figno}%
\figinsert\figin{\centerline{#3}}\medskip\centerline{\vbox{\baselineskip12pt
\advance\hsize by -1truein\center\footnotesize{  Fig.~\the\figno.} #2}}
\bigskip\endinsert\global\advance\figno by1}
\def\endinsert{}

\begin{document}

\title{\large{\bf Isospin Fluctuations from Multiple Domains of
Disoriented Chiral Condensate}}

\author{James Hormuzdiar\thanks{james.hormuzdiar@yale.edu}~
\\ Department of Physics, Yale University, New Haven, CT 06520-8120 
\\ \\
Stephen D.H.~Hsu\thanks{hsu@duende.uoregon.edu}
\\ Department of Physics, University of Oregon, Eugene, 
OR 97403-5203
\\
}

\date{February, 1998}

\maketitle

\begin{picture}(0,0)(0,0)
\put(350,350){OITS-646}
\put(350,330){YCTP-P2-98}
\end{picture}
\vspace{-24pt}

\begin{abstract}
We describe some detailed numerical simulations of Disoriented
Chiral Condensates (DCCs), using the chiral lagrangian as a 
controlled long-wavelength description. We focus on the possibility
of multiple, independently coherent domains, and investigate the 
degree to which the DCC signal is attenuated. As an intermediate step 
in our analysis we compute the expected number of detector events in each
isospin and momentum channel for a given
asymptotic classical field configuration. We find that for
sufficiently large initial field strengths, the non-linear interactions
between domains become important and can lead to a randomization of 
isospin orientations. Nevertheless, we argue that viable signals 
exist for DCC detection, even in the case of multiple domains and
strong domain-domain interactions. We briefly discuss some long-lived
`pseudo-bound state' configurations which arise at large field strengths
and might be observable in HBT correlations.

\end{abstract}

\newpage
\section{Introduction}

One of the most exciting possibilities at RHIC is that collisions
of heavy nuclei might lead to coherent, pion field configurations 
which are produced by the disorientation of the chiral condensate (DCC)
\cite{Anselm88,Anselm91,Blaizot:PRD46,
Rajagopal:NPB399,Bjorken} (for recent review articles, see 
\cite{Rajagopal:QGP2,BK:96}).
Such coherent configurations could lead to large, characteristic fluctuations
in the number and type of pions measured in detectors. 
Unfortunately,
the evolution of quantum fields in the aftermath of the heavy ion
collision is a difficult problem, involving both strong coupling effects
and non-equilibrium statistical mechanics, and remains to be solved
despite much theoretical effort. 
Most previous work has been performed in 
the context of the linear sigma model, sometimes using an approximation like
the $1/N$ expansion to control the quantum corrections  
\cite{Rajagopal:NPB404}--\cite{JR:PRL}.
However, the relationship between linear sigma model dynamics and real
QCD dynamics is unclear.

Our approach here will be somewhat different.
We adopt an agnostic approach to the {\it formation} of DCCs, and
simply try to characterize their evolution at subsequent times, with
an eye toward experimental signatures.
We make a detailed investigation of the phenomenology of DCCs using
the classical field equations derived from the chiral lagrangian (non-linear
sigma model)
to describe their evolution from initial domain(s) to the final state
which arrives in the detector. The field equations of the
chiral lagrangian can be organized
in a momentum expansion, and therefore yield a controlled approximation
for sufficiently slowly-varying pionic configurations. 
While this approach has no advantages for the formation problem, it does
provide a way of simulating, in a controlled fashion, the subsequent
evolution of domains once formed.
Indeed, quantum corrections (loop effects)
are suppressed as long as the configurations simulated are sufficiently
``soft'' relative to the scale $4 \pi F_{\pi} ~\simeq 1 \, {\rm GeV}$.
We verify that this treatment
is self-consistent: initially soft configurations do not lead to the
build up of hard sub-configurations which cannot be treated within the
chiral lagrangian.

We will be particularly interested
in the possibility of multiple coherent domains, since it seems unlikely
that a single large domain would remain in the wake of the collision
\cite{Cooper,Boyanovsky:PRD51,Kluger}.
Therefore, we simulate the interactions between domains with different
initial isospin orientations. For sufficiently weak fields, interactions
can be neglected, leading to particle distributions which are simply
linear superpositions of those expected from each separate domain.
In this case the effect of $N$ multiple domains is mainly ``statistical'',
leading to a narrower width (proportional to $1 / \sqrt{N}$) in 
fluctuations. 
However, at sufficiently large field strengths, which we quantify below (but
which are well within the roughly expected range),
the interactions are significant and have a strong effect on the
field configurations, even if the number of domains is small.
Our results are discussed in detail below, however the general
observation one can make is that non-linearities can degrade the
DCC signal, leading to a somewhat narrower distribution in fluctuations
of the charge ratios. Despite this, we find that promising signals
of multiple DCC domain formation remain even in the non-linear cases.
We also briefly discuss some interesting `pseudo-bound state' (PBS) 
configurations
which appear at large field strengths. These configurations have particularly
long lifetimes and might be observable in HBT correlations at RHIC.

The organization of this paper is as follows. In section 2 we review
the chiral lagrangian approximation to low-energy QCD, and derive the
corresponding classical field equations. In section 3 we discuss various
aspects of multiple domains, including the straightforward statistical
analysis. In section 4 we describe our simulation and our numerical
results, including details of the non-linear behavior. 
In section 5 we present our conclusions. Appendices A, B and C
contain some useful formulae, as well as a description of our ``detector''
subroutine and of how isospin rotations of our configurations were
performed. In appendix D we give additional details concerning the simulations
and the parameters used.

\section{Chiral Lagrangian Description}

The chiral lagrangian is the most general effective lagrangian consistent
with the symmetries of low energy QCD \cite{CL}. As such, it accurately
describes the soft dynamics of pseudo-Goldstone mesons, including the virtual
effects of heavier particles such as baryons, the rho meson, etc. which
have already been integrated out.
We can classify the
possible terms by the number of derivatives, with the first term given by
\beq
\label{chiralL}
{\cal L}
= {F_\pi ^2 \over 4}tr(\partial^\mu \Sigma^\dagger \partial_\mu \Sigma)
\eeq
where
\beq
\Sigma = e^{i\pi \cdot \vec \tau /F_\pi} ~.
\eeq
The effects of higher order terms are proportional to powers of the
momentum, as are any quantum loop effects calculated with (\ref{chiralL}).
This implies that for sufficiently soft configurations, the classical 
field equations derived from (\ref{chiralL}) describe the full quantum
mechanical evolution. For this reason, we
believe that the chiral lagrangian is superior to the usual linear sigma
model for studying the dynamics of DCC's, at least after formation. Our
approximation relies on a momentum expansion, rather than $1/N$ or
Hartree-Fock.

We define a four component real field $\phi=(\sigma, \vec \phi)$ such
that $F_\pi \Sigma = \sigma I+i\vec \tau \cdot \vec \phi$ with the
condition $\phi^2 = F_\pi^2$ everywhere.  It is clear that we can do this,
since
\beq
e^{i\vec \tau \cdot \vec A} = \cos(|\vec A|)I
+i{\vec \tau \cdot A \over |\vec A|}
\sin (|\vec A|) ~.
\eeq
In terms of $\phi$, we have
\beq
\label{Lphi}
{\cal L} = {1\over 2} \partial^\mu \phi \cdot \partial_\mu \phi~~,
\eeq 
with
the condition $\phi^2 = F_\pi^2$.
 
Rather than deal with the complicated constrained equations which result
from (\ref{Lphi}),
we relax the condition on $\phi^2$
slightly by replacing it with a potential term in the lagrangian whose
minimum falls at $\phi^2=F_\pi^2$. This yields the linear sigma model 
with potential
\beq
V(\phi) = {\lambda\over 4}[{\phi}^2 - F^2_\pi]^2,
\eeq
and lagrangian
\beq
{\cal L} = {1\over 2} {\partial}^\mu \phi \cdot {\partial}_\mu \phi - V(\phi)
={1\over 2} {\partial}^\mu \phi \cdot {\partial}_\mu \phi -
{\lambda\over 4}[{\phi}^2 - F^2_\pi]^2~~.
\eeq
The limiting case $\lambda \to \infty$ yields the original
non-linear sigma model.

In reality, the chiral symmetry is slightly broken by bare quark masses,
resulting in non-zero pion masses.  
Taking this into account yields
\beq
{\cal L} = {1\over 2} {\partial}^\mu \phi \cdot {\partial}_\mu \phi -
{\lambda\over 4}[{\phi}^2 - F^2_\pi]^2
-{m^2_\pi\over 2}(\phi^2-2F_\pi \sigma + F_\pi^2) .
\eeq
This lagrangian yields the classical equations of motion
\begin{eqnarray}
\partial^2 \vec \phi &=& -[\lambda(\phi^2-F_\pi^2)+m_\pi^2]\vec \phi \\
\partial^2 \sigma &=& -[\lambda(\phi^2-F_\pi^2)+m_\pi^2]\sigma+m_\pi^2 F_\pi~~.
\end{eqnarray}

In the non-linear sigma model limit we wish to consider, $\sigma$ is just 
$F_\pi$ plus a very small
quantity. Because floating point variables do not deal well with this 
arrangement, it is more useful to redefine
$\sigma \rightarrow \tilde \sigma + F_\pi$, 
with equations of motion
\begin{eqnarray}
\label{eq:ofmotion}
\partial^2 \vec \phi &=& -[\lambda(\tilde \phi^2+2F_\pi \tilde \sigma)
+m_\pi^2]\vec \phi \\
\nonumber
\partial^2 \tilde \sigma &=&
-[\lambda(\tilde \phi^2+2F_\pi \tilde \sigma)+m_\pi^2]\tilde \sigma
-\lambda(\tilde \phi^2+2F_\pi \tilde \sigma)F_\pi. 
\end{eqnarray}

\section{Multiple Domains}
\label{sec-multdom}

A telltale sign of DCC formation would be large fluctuations in pion isospin abundances.
Under incoherent production of hundreds or possibly thousands  
of pions, one would expect nearly equal numbers of each type to be emitted,
whereas with coherent DCC domains, the abundances
would be a function of the domain isospin orientation.  
If one domain happened to point in the $\pi_0$ direction,
then each of the outgoing pions would be a $\pi_0$, 
an improbable occurrence under incoherent production.
In the case of a single, coherent domain, the exact probability
distribution for 
\beq
f \equiv { N_{\pi^0} \over N_{\pi^0} + N_{\pi^+} + N_{\pi^-} }
\eeq
is easily seen to be $dP(f) = df / 2 \sqrt f$ .

\begin{figure}[htb]
\center{
\leavevmode
\epsfbox{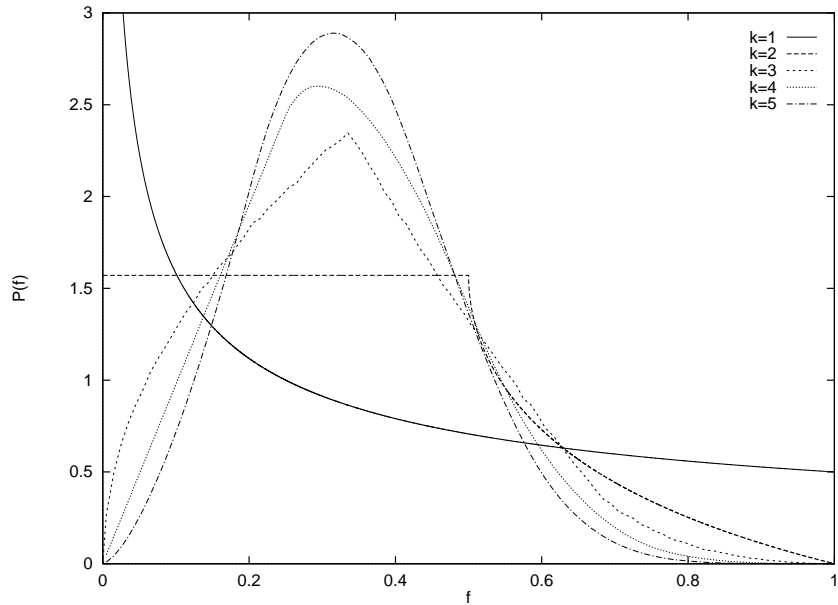}
\caption{Probability distribution of $f$ 
[$P(f)$]for varying k} \label{fig:p(f)_low_strength}
}
\end{figure}

The case of k multiple domains is more complicated, but in the 
low field strength 
approximation it can be simplified by the neglect of interactions,
so that the distribution of pions is the sum of the distribution from
each single domain.  If the size of each domain is taken to be
roughly the same, and hence total particle output 
$N = N_{\pi^0} + N_{\pi^+} + N_{\pi^-}$ 
does not differ from domain to domain, then
the total $f$ is simply the average of that for the individual domains,
\beq
f_{total} = {\sum N_{\pi^0}^j \over \sum N_j} = {\sum N_{\pi^0}^j \over k N}
= {1 \over k} \sum f^j~~.
\eeq  
As was also shown in \cite{stat}, the distribution for this total
$f$ can now be computed as a function of k.  
Analytical results are difficult to obtain, but by 
randomly generating domain orientations 
(with a flat probability distribution on the three sphere), 
the probability 
distribution curves
can be calculated numerically.  By generating large 
quantities of
samples, the error bars in the distribution curves were 
lowered enough to be negligible in the 
graphs, approximating the limiting case of a continuum 
probability distribution.
Figure~\ref{fig:p(f)_low_strength} shows the behavior 
of $P(f)$ for increasing values of k.

\epsfysize=8.0 cm

The general behavior is a result of the central limit theorem, 
which tells us that as 
statistical sampling is 
performed repeatedly, the average value of the
resultant variable stays the same, while its standard 
deviation is proportional to $1/\sqrt k$.  
The proportionality constant is easy to fix by calculating
the standard deviation for the k = 1 case, whose probability 
distribution was
given above as $dP(f) = df / 2 \sqrt{f}$.  
This yields 
\begin{equation}
\sigma_{\rm coherent} \equiv \sqrt {\langle f^2 \rangle -  
\langle {f} \rangle^2} 
= {2 \over 3 \sqrt{5k}}~~.  \label{eq:coherent}
\end{equation}

We can repeat this calculation for the incoherent case with N particles
emitted, using the probability distribution for 
single particle production, 
$$P(f) = {2 \over 3} \delta(f) + {1 \over 3}\delta(f-1)   .$$
In this case
\beq
\sigma_{\rm incoherent} = {1 \over  3} \sqrt {2 \over N}~~. \label{eq:incoherent}
\eeq
While we do not know precisely how many domains (if any!) are likely
to be formed in a typical collision, it is difficult to imagine that
$k$ is much larger than of order 10 or so, simply due to the restricted
geometry of the collision region. 
Therefore, $N$ is likely to be large compared to $k$ and
there is still a significant qualitative difference between the 
distribution of results
from incoherent and coherent production. The overall level of fluctuations
in $f$ should be larger in the coherent case.
Whether this difference could be observed at RHIC
depends of course on the probability of DCC formation per collision.

The coherent analysis would be modified if some 
fraction $a$ of the outgoing pions were 
assumed to be produced by 
incoherent background processes.  
For $(N'_{\pi_0}, N'_{\pi_+}, N'_{\pi_-})$ coherently produced and 
$(N''_{\pi_0}, N''_{\pi_+}, N''_{\pi_-})$ incoherently produced pions, 
(where $N''_{total} = a (N'_{total} + 
N''_{total})$), the value of $f$ is (neglecting effects of order 
$1 / \sqrt{ N''}$)

\beq
f ~=~ {N'_{\pi_0} + N''_{\pi_0}\over N'_{total} + N''_{total}} ~=~ 
(1-a) f_{\rm{non-corrupted}} + {1\over 3}a~~.
\eeq
Using this the standard deviation of $f$ becomes 
\beq
\sigma_{\rm coherent} 
= (1-a){2 \over 3 \sqrt{5k}}~~.
\eeq
Hence incoherent corruption narrows the spread in $f$ around 1/3.

In the above discussion we ignored the interactions between the coherent
configurations. Clearly this is justified for sufficiently weak fields, but
when the pion field strengths are large the interactions
between different domains can be significant. It is well known that 
the solution of classical field equations is equivalent to the 
resummation of all tree graphs in perturbation theory. 
Nonlinearities in the classical field evolution therefore represent large
rescattering effects due to tree-level vertices. As mentioned previously,
all loop graphs in chiral perturbation theory contain extra powers of
the external momentum squared, and are hence suppressed.
We expect rescattering
to be important when the pion number densities are sufficiently large.
The effect of the interactions is not clear {\it a priori}, although
one might argue that they tend to randomize the isospin orientations
and lead to smaller, not larger, fluctuations in $f$ for a fixed number
of domains.
To gain some insight on this question, we performed numerical
simulations of domain-domain interactions using the low energy 
effective field
theory described in the previous section.

In our simulations we allowed the field strengths to vary in the range
of $\vert \vec{\pi} \vert \sim F_{\pi}$. Due to our lack of control over
the formation process, we are unable to compute with any precision
the expected number or energy density of pions in the DCC. However, a simple 
estimate based on a critical temperature of chiral symmetry restoration
$T_c \simeq 200 \, {\rm MeV}$, and the assumption that the energy density
of pions in the DCC is roughly that of the hadron gas at $T_c$, yields
field strengths as large as $(4-5)~ F_{\pi}$. As we will see below, 
the evolution of multiple domains is significantly non-linear at field
strengths which are as small as $\vert \vec{\pi} \vert \simeq 1.4  F_{\pi}$.

\section{Simulation and Results}

We performed simulations of domain--domain interactions in a cylindrical
geometry with axial symmetry. The domains themselves are cylinders placed
end to end along the beam axis, with domain walls in between.
This is of course an idealization of the actual configuration, chosen for 
simplicity and to make our simulations tractable.

\subsection{The Difference Equation}

A computer simulation written in C++ was used 
to evolve eq.~(\ref{eq:ofmotion}).  
Imposing axial symmetry, the left side of the equation 
becomes
\beq
\partial^2 \phi =
{\partial^2 \phi\over\partial t^2}-
{\partial^2 \phi\over\partial z^2}
-{1\over r}{\partial \over\partial r}[r{\partial \phi\over\partial r}]
={\partial^2 \phi\over\partial t^2}-
{\partial^2 \phi\over\partial z^2}
-{1\over r}{\partial \phi\over\partial r}
-{\partial^2 \phi\over\partial r^2}~~.
\eeq
Defining a discrete function
\beq
\phi_{i,j,k} \equiv \phi(i \Delta t, j \Delta r,k \Delta z),
\eeq 
and using the following finite-difference
approximations 
\begin{eqnarray}
{\partial \phi\over \partial q} &\simeq& {\phi_{i+1}-\phi_{i-1}\over 2\Delta q} \\
{\partial^2 \phi\over \partial q^2} &\simeq& {\phi_{i+1}-2 \phi_i+\phi_{i-1}\over 
\Delta q^2} ~~,
\end{eqnarray}
the equations of motion become
\begin{eqnarray}
\phi_{i+1, j, k} &=& 2 (1-2{\Delta t^2\over\Delta z^2})\phi_{i, j, k} 
+ {\Delta t^2\over\Delta z^2}[ (1+{1\over 2j})\phi_{i, j+1, k} \nopagebreak
\\
\nopagebreak
\nonumber &+&~
\nopagebreak
(1-{1\over 2j})\phi_{i, j-1, k}+F_{i, j, k+1}+F_{i, j, k-1}]
-F_{i-1, j, k}+\Delta t^2 G(\phi_{i, j, k})~~.
\end{eqnarray}
(Here $\Delta r = \Delta z$, and $G(\phi_{i, j, k})$ is the expression on 
the right hand side of the original 
equation).

By storing the two previous time slices of the field 
($\phi_{i, j, k}$ and $\phi_{i-1, j, k}$),
the newer value ($\phi_{i+1, j, k}$) can be calculated, 
allowing for the time evolution of the fields
given an initial configuration and its time derivative. 

\subsection{Initial Configurations}

An initial configuration is described by its number of domains k, 
their positions, sizes, field strengths, and 
isospin orientations.  For simplicity we assume the size and 
strength of each domain is the same (owing to a 
similarity in formation processes).  We also choose the 
field configurations of the first two timeslices to be the 
same. As a final simplifying assumption we have ignored the
initial expansion velocities of the domains. Such velocities would
probably {\it decrease} the nonlinear effects, since the pions would
disperse more rapidly to low densities.

Distinct fixed values of k are treated differently.

\bigskip
\bf
\leftline{k=1}
\rm

\nopagebreak

Here the initial conditions for each run are characterized by one vector $\vec v$ 
representing the orientation and strength of 
the single domain in isospace.

Varying the field strength in the k=1 case yields no change in the distribution curve for $f$.  
It can be easily shown (e.g., using the 
results in appendix C) that the final values of $N_{\pi_0}$, 
$N_{\pi_+}$, and $N_{\pi_-}$ are proportional to the 
square of the three components of $\vec v$:
\beq
(N_{\pi_0}, N_{\pi_+}, N_{\pi_-}) \propto (v_{\pi_0}^2, v_{\pi_+}^2, v_{\pi_-}^2)
   ~~.
\eeq
The proportionality constant cancels out in the definition of $f$, 
leaving it and its statistics invariant to the 
initial strength.

\bigskip
\bf
\leftline{k=2}
\rm

\nopagebreak

The initial configurations for the k=2 case can be 
parameterized by two isospin vectors $\vec v_1$ and $\vec v_2$.  
To generate the $P(f)$ curves, the field strength of the 
two domains was fixed at the given value 
$|\vec v_1| = |\vec v_2| = s$, multiple runs with random 
orientations were performed, and the resulting $f$'s were 
binned, much like in the weakly interacting case.

Each run requires tens of minutes, hence it is
too time consuming to perform the $O(10^9)$ simulations 
needed to smooth the probability curves.  Instead, the 
rotational techniques developed in appendix C were exploited, 
and the simulation was only run for $\vec v_1$ and 
$\vec v_2$ orientations in the $\pi_0-\pi_+$ plane, 
symmetric around the $\pi_0$ axis.
Rotating this representative class yields all possible 
initial configurations.

The actual functional shape of the initial field configurations is specified by 
a magnitude
and angular dependence. The magnitude is given by
\beq
|\vec \pi (r, z)|= s * g(|z|)g(r)~,
\eeq
with $g(x) = {1 / \left( 1+e^{k_1 (x-a)} \right) }$. This function
(for appropriate $k_1$'s) is nearly constant for $x < a$, drops of suddenly at 
$x \simeq a$, then stays approximately 0 for $x > a$.
The angular dependence, measured in the $\pi_0-\pi_+$ plane from the $\pi_0$ axis, is
\beq
\angle \vec \pi (r, z) = {2 ~ \alpha \over \pi} ~ \arctan (k_2 ~ z),
\eeq
which plateaus to $+\alpha$ for $z>0$ and $-\alpha$ for $z<0$.
These equations describe a cylinder centered at $z=r=0$ subdivided into a 
$z>0$ part with isospin in the $+\alpha$ 
direction and a $z<0$ part with isospin in the $-\alpha$ direction.  
Values of $k_1$, $k_2$, and $a$ were chosen to 
fix the physical dimensions. Each isospin region has a  
height and diameter of $10 \, {\rm fm}$, with a 'skin' thickness of 
$1 \, {\rm fm}$. (The
'skin' is defined as the region where 
the field amplitude drops from $90\%$ to $10\%$ of its maximum value.)
Additional details of the 
computational techniques used in the simulations can be found in appendix D.

\epsfysize=10.4 cm
\begin{figure}[htb]
\center{
\leavevmode
\epsfbox{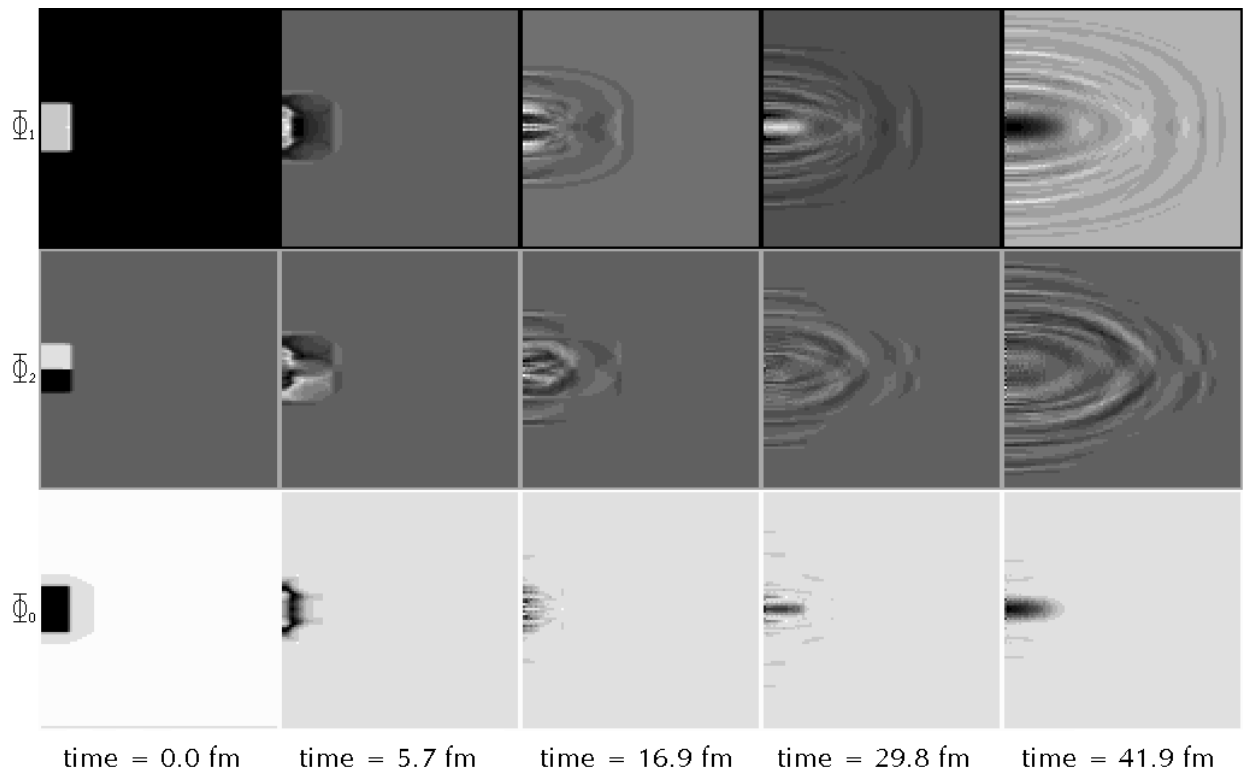}
\caption{Numerical evolution of the fields} \label{fields}
}
\end{figure}


Figure~\ref{fields} gives sample time snapshots in the evolution of the pion fields, 
where light white (dark black) corresponds to the maximum (minimum) value on the field in the 
given snapshot.
Runs were continued until the Klein-Gordon Number Operator converged, signaling that the 
field had spread out far enough that the nonlinear terms had become negligible.

\begin{figure}[htb]
\center{
\leavevmode
\epsfbox{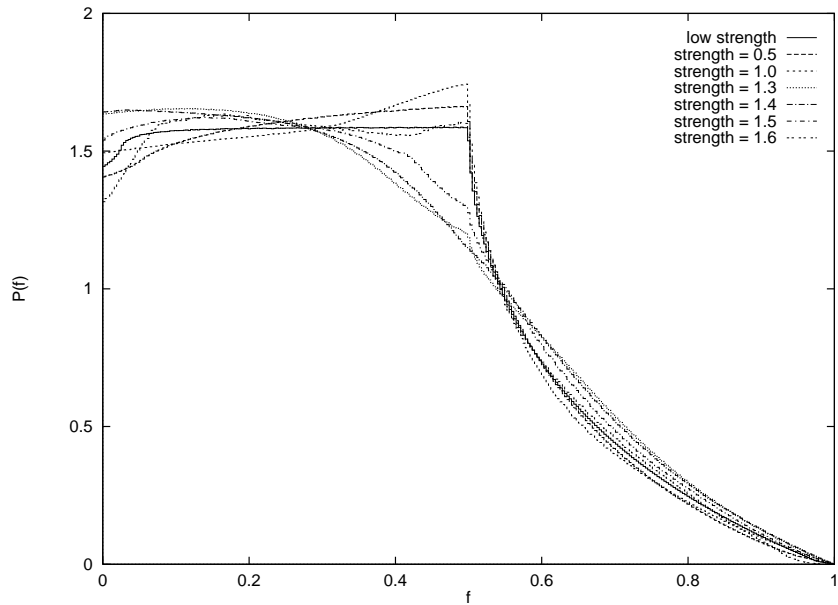}
\caption{Probability distribution of $f$ [$P(f)$] for varying field strength} 
\label{fig:p(f)forvaryk}
}
\end{figure}

The progression
of $P(f)$ for various 
initial field strengths was constructed and is shown in figure~\ref{fig:p(f)forvaryk}.
The dotted line shows the low field strength approximation.  
Successive curves are then shown for initial
field strengths of 
$0.01 F_\pi$, $0.5 F_\pi$, $1.0 F_\pi$, $1.3 F_\pi$, $1.4 F_\pi$, $1.5 F_\pi$, and $1.6 F_\pi$.
The trend towards a more randomized probability distribution is apparent, although
the width of the distribution and the area under the tails (in particular, $f \simeq ~1$) is
not changed significantly. One way to understand this is to note that in order to get an
event with $f$ close to unity, the initial domains must all be oriented in the same direction.
This is of course less and less probable as the number of domains $k$ gets larger, but once
the domains are formed in this way the common orientation suppresses their interactions --
in other words they act like one big domain.
Thus we expect the number of $f \simeq ~1$ events to be preserved 
even at large field strengths.

\nopagebreak

\epsfysize=8.0 cm

\bigskip
\bf
\leftline{k $>$ 2}
\rm

Computational power limited our ability do do an 
exhaustive analysis of the k$>$2 case.  
For the k=3 case, initial 
configurations are parameterized by the three unit vectors and an 
overall strength.  Even exploiting the rotational 
symmetry and fixing field strength, three free parameters 
remain.  Building a representative class of configurations, 
sampling each variable at S values would 
would need at least $S^3$ computer runs.  Even the marginally 
acceptable value of $S = 10$ would 
require 
1000 runs, which at greater than 2 hours per run on our DEC
Alpha Workstation would require almost 3 months.  
Although more powerful computers could be assigned 
to this task, we feel that the important qualitative 
features of the domain to domain interaction are already 
featured in the k=2 case.  This intuition is 
based in part on the (explicitly confirmed) observation 
that when two domains are separated from each 
other at larger distances, interactions become quickly 
negligible. Thus, in the case of k domains 
arranged colinearly (as required by the axial symmetry) only the nearest 
neighbor domain walls would interact significantly.  Of course this single domain wall was 
well studied in the k=2 case.

\subsection{Nonlinearities and Pseudo-Bound States}

\label {nonlin}

\begin{figure}[htb]
\center{
\leavevmode
\epsfbox{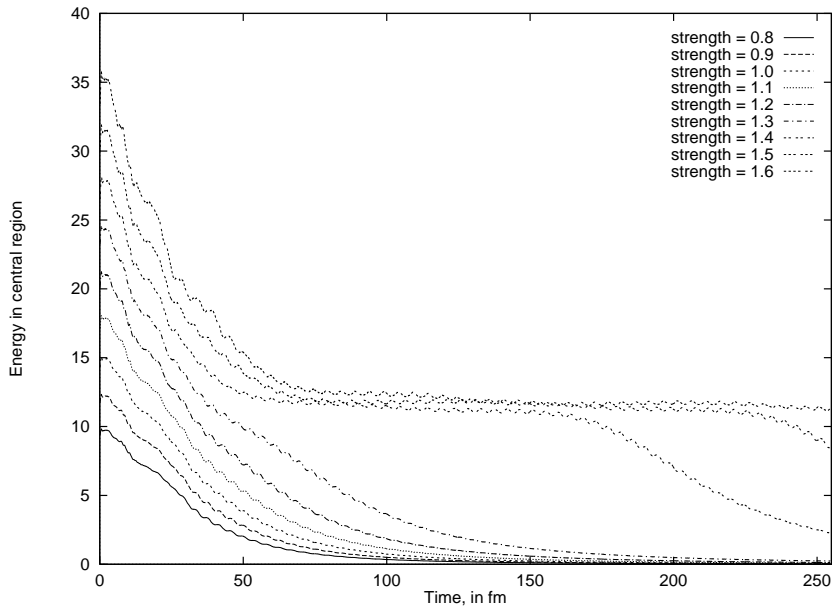}
\caption{Central energy vs. time} \label{pseudo-bound-state}
}
\end{figure}

Increasing the strength of the field configurations qualitatively changes the evolution of the 
fields.  We carefully examined the runs to determine the critical
amplitude at which nonlinearities set in.  
In these observations a surprising behavior was noticed.  For values of initial 
strengths between $0$ and $1.2 F_\pi$, evolution proceeded similarly 
to the non-interacting  case. However, for slightly higher field strengths we
observed that a large
fraction of the energy bunched up in the central region 
and held together for long periods of time.
Figure~\ref{pseudo-bound-state} shows the energy in a $10 \, {\rm fm}$
spherical region centered around the origin vs. time.  As can be
seen in these graphs, once the threshhold has been crossed a pseudo-bound
state forms.

\epsfysize=8.0 cm

This phenomenon was found to be robust: it was replicated 
in runs in which the initial conditions were varied. We also studied the evolution
of spherically symmetric configurations using  
separately written code.  Again, once a certain initial strength was crossed, the
configuration held together in a pseudo-bound state for much longer periods of time.  
It is in fact the long evolution times required for the configuration to completely disperse
that limited our ability to compute the 
probability distribution in $f$ beyond the $1.6 F_\pi$ case.

These much longer dispersion times are potentially observable in HBT (Hanbury Brown-
Twiss) correlations\footnote{We thank K. Rajagopal and D. Rischke for pointing this
out to us.}. (For a recent review of these techniques, see \cite{HBT}.)  
Pseudo-bound state dispersion times as long as  $10^{-22}s$ (versus $10^{-24}s$ for the 
noninteracting case) were observed, and we have every indication that longer 
lifetimes are possible. We intend to report in more detail on PBS dynamics in
future work.

We note that although the configurations evolved very differently for higher 
field strengths, the $P(f)$ curves were only slightly modified.  
This fact leads us to assert that, at least up to the strengths that we were able to study, 
the noninteracting case remained a good approximation for the full evolution.

\section {Signals}

In this section we discuss two signals which can be used to detect or exclude DCC production.
We focus on simple hypotheses with specific probabilities $p_k$ for producing $k$ domains of some 
fixed characteristic size.

\subsection{Width Measurements}
Purely incoherent pion production should yield a spread in values of $f$ equal to 
$\sigma = {1\over 3}\sqrt{2\over N}$ (equation~\ref{eq:incoherent}).
Any significant excess of this value would be strong evidence of DCC formation.  Although the 
range of probable RHIC 
pion yields is
presently unknown, experience from previous heavy ion collisions would suggest values in 
the hundreds or thousands, 
yielding $\sigma = 0.01$ to $0.05$, a range that would be surpassed for up to k=35 
domains (equation~\ref{eq:coherent}).

In reality, k domain DCCs would be expected to form in only 
some fraction $p_k$ of the events, yielding a smaller
value of $\sigma$.  Data can place upper bounds on each value 
of $p_k$ from the following analysis.  Assuming that 
only k domain DCC's and incoherent events form (for some given k), 
the probability distribution curve for $f$ becomes

\beq
P(f) = p_k~P_{\rm{k-domains}}(f) + (1-p_k)~P_{\rm{incoherent}}(f)~~,
\eeq
hence
\begin {eqnarray}
\sigma_{\rm{measured}}^2 = \int df (f-<f>)^2 ~ [p_k ~ 
P_{\rm{k-domains}}(f) + (1-p_k) ~ P_{\rm{incoherent}}(f)]
\\
\nonumber
= p_k ~ \sigma_{\rm{coherent}}^2 + (1-p_k) ~ \sigma_{\rm{incoherent}}^2~~.
\end {eqnarray}

Using equations~\ref{eq:coherent} and~\ref{eq:incoherent}, 
and reintroducing incoherent background corruption discussed in section~\ref{sec-multdom}, 
we can obtain an upper limit of $p_k$

\beq
p_k ~ \leq ~{5\over 2}k~ \left[ {9N\sigma_m^2 - 2\over 2(1-a)^2 N-5k} \right]
\eeq

\begin{figure}[htb]
\center{
\leavevmode
\epsfbox{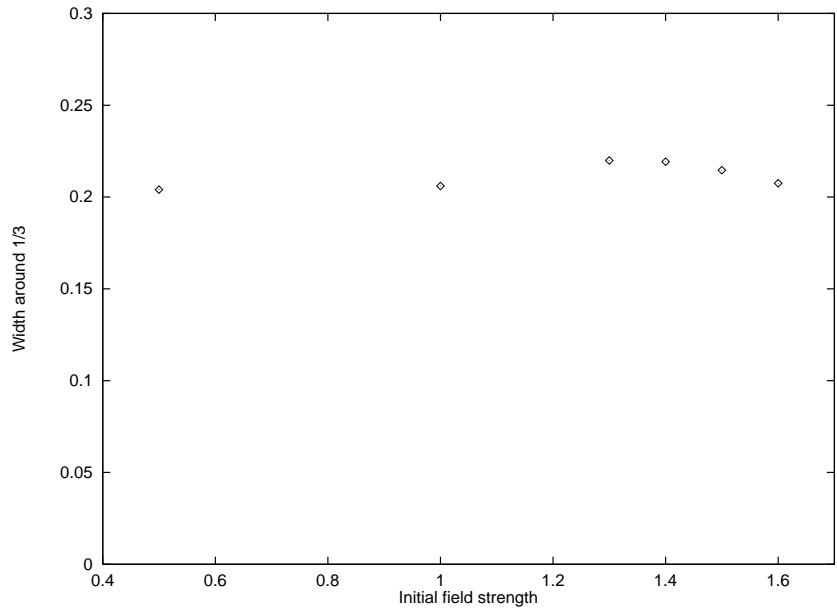}
\caption{Standard deviation vs. initial field strength} \label{fig:disp}
}
\end{figure}

Figure~\ref{fig:disp} shows a plot of numerically calculated values of 
width vs. initial field strength (for 
the $k=2$ case).  Despite the level of nonlinear activity that occurs at higher strengths,
the width does not change more than $10\%$.  Up to the range of simulated cases, the noninteracting
case is a good approximation. For higher strengths, there is probably a trend toward somewhat
smaller widths.

\epsfysize=8.0 cm

\subsection{Counting Rare Events}

Another feature of incoherent production is that the distribution of 
$f$ peaks sharply at 1/3, and is nearly zero elsewhere, yielding an 
exponentially small tail of events where $f$ is close to either $0$ or $1$.
From
elementary statistics the distribution of $f$ is

\beq
P(f=i/N)=
\left( \begin{array}{c}
N \\ i
\end{array}  \right) 
\left( {1\over 3} \right)^i ~ \left( {2\over 3} \right) ^{N-i}.
\eeq
where $i$ is restricted to being an integer from $0$ to $N$.  Setting N=100 and 
integrating over the range $0.6\leq f\leq 1$ yields the small 
value of $4.3\times 10^{-8}$. Thus, given (for instance)
$10^7$ events, there is still only a 43\% chance that 
$f\geq .6$ be measured even 
once.

A similar anaylsis for 
the coherent case was performed.  For instance, with one 
domain, $P(f)={1\over 2 \sqrt f}$, for which 23\% of the events lie above $f=0.6$
(a much larger $2.3\times 10^6$ of $10^7$ assumed events).  
Now, assuming that only the incoherent and one domain coherent cases
contribute, the fraction of rare events $t$ would be
\beq
t = (1-p_1)~t' + p_1~t''
\eeq 
where $t'$ is the number of rare events expected from the incoherent
case, $t''$ is the number from the coherent case, 
and $p_1$ is an element in the set of $p_k$'s introduced in the previous subsection.  
Presumably more than just these two cases would exist, so this equation sets an upper limit on $p_1$
\beq
p_1\leq {t-t'\over t''-t'}~~.
\eeq

In this manner, using the probability curves of $f$ calculated in previous sections, data for other
values of k were calculated.  Table~\ref{fig:rare_events} shows the expected fraction  
of tail events for given domain configurations (varying k, fraction 
of incoherent corruption, and field strengths for k=2).  Also included are upper bound 
expressions for the $p_k$'s. 

\begin{table}
\begin{tabular}{|c|c|c|} \hline

Number of domains k & Fraction of events & Upper bound on $p_k(\leq {t-t'\over t''-t'})$ \\

 & with $f>0.6$ & \\ \hline \hline

$k=1,~ {\rm all~field~strengths}$ & $2.3\times 10^{-1}$ & ${4.3~ t -1.9\times 10^{-7}}$ \\ \hline
$k=1,~ {\rm all ~field~ strengths}~^{\dagger}$ & $1.8\times 10^{-1}$ & ${5.4~ t-2.3\times 10^{-7}}$ \\ \hline
$k=2,~ {\rm Low ~field ~strengths}$ & $1.2\times 10^{-1}$ & ${8.3~ t-3.6\times 10^{-7}}$ \\ \hline
$k=2,~ {\rm Low ~field ~strengths}~^{\dagger}$ & $7.4\times 10^{-2}$ & ${14~ t-5.8\times 10^{-7}}$ \\ \hline

$k=2,~ {\rm High~field~strengths}$ & $1.1\times 10^{-1}$ & ${9.1~ t-3.9\times 10^{-7}}$ \\ \hline
$k=2, ~{\rm High~field~strengths}~^{\dagger}$& $7.1\times 10^{-2}$ & ${14.1~ t-6.1\times 10^{-7}}$ \\ \hline
$k=4$ & $5.0\times 10^{-2}$ & ${20~ t-8.6\times 10^{-7}}$ \\ \hline
$k=4~^{\dagger}$ & $1.8\times 10^{-2}$ & ${56~ t-2.4\times 10^{-6}}$ \\ \hline
$k=8$ & $8.4\times 10^{-3}$ & ${120~ t-5.1\times 10^{-6}}$ \\ \hline
$k=8~^{\dagger}$ & $1.4\times 10^{-3}$ & ${710~ t-3.1\times 10^{-5}}$ \\ \hline
$k=16$ & $3.4\times 10^{-4}$ & ${2900~ t-1.3\times 10^{-4}}$ \\ \hline
$k=16~^{\dagger}$ & $1.1\times 10^{-5}$ & ${9.1\times 10^{4}~ t-3.9\times 10^{-3}}$ \\ \hline
$k=32$ & $7.7\times 10^{-7}$ & ${1.3\times 10^{6}~ t-5.9\times 10^{-2}}$ \\ \hline
$k=32~^{\dagger}$ & $8.8\times 10^{-10}$ & ${2.4\times 10^{7}~ t-1.0}$ \\ \hline
\end{tabular}

($^{\dagger}$ = with 20\% incoherent background) 

\caption{Rare events}\label{fig:rare_events}
\end{table}

\begin{figure}[htb]
\center{
\leavevmode
\epsfbox{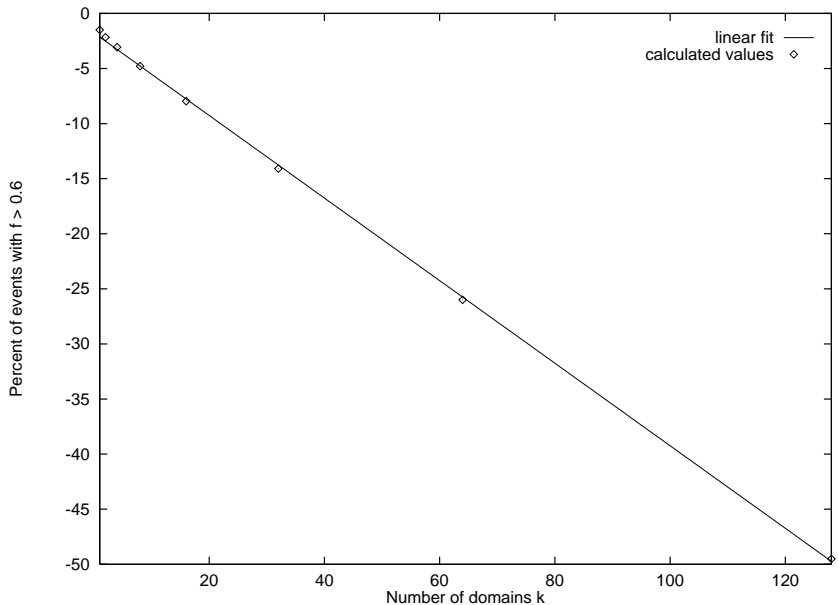}
\caption{Tail events versus k} \label{fig:tail_vs_k}
}
\end{figure}

From this data, we can conclude that the fraction of events lying in 
the tail falls off exponentially with $k$.  Figure~\ref{fig:tail_vs_k} shows a fit to the form 

\beq
P_{\rm{tail}} = A e^{-B k}~~.
\eeq
with fitted values $A=0.147$ and $B=0.370$.  From this, 
it is estimated that k could increase to 40 before the number 
of tail events would be comparable to the incoherent case.  Similarly, with 20\% incoherent 
background included, $A=0.20$, $B=0.61$, and k can still 
increase to around 25 before the number of tail 
events would be comparable to the purely incoherent case. 

\epsfysize=8.0 cm

\section{Conclusion}

We have conducted a detailed numerical investigation of multiple
domains of disoriented chiral condensate. Although the dynamics
of domain formation in the wake of a heavy ion collision is far 
from completely understood, it is plausible that multiple domains
are more likely to occur than a single coherent domain. We have 
concentrated on the evolution of multiple domains after formation,
using the chiral lagrangian as a long-wavelength approximation for
the dynamics. Our simulations follow the pion fields out to asymptotic
distances, where we can characterize the number
of particles per isospin and momentum bin that would be detected in
an experiment. In choosing the initial conditions for our configurations,
we opted for simplicity in neglecting important factors such as the
expansion velocities. Given a particular dynamical model of the
formation process, this could easily be incorporated into our simulations.

For small amplitudes, corresponding to low number densities of pions
in the semi-classical fields, a simple statistical analysis suffices
to characterize the isospin fluctuations of multiple domains (see figure
1). However, as the field strengths are increased, we see ample evidence
of nonlinear behavior and significant interaction between domains. These
nonlinear effects are already present for field strengths of order 
$\sim F_{\pi}$. For the case of two domains ($k = 2$), we were able to 
compute the effects of interactions on the probability distribution of isospin
fluctuations (figure 2). While the shape of the curve is changed somewhat
from the weak-field case, the probability of rare events (in particular, where f
is close to unity) and the width of the distribution 
are not strongly affected. We suspect that the rare events with large f will
prove to be the most robust signal of DCCs, surviving even 
multiple domain formation and large field strengths.
Some interesting phenomena
were observed in the evolution with high initial field
strengths. Relatively long-lived `breather' or `pseudo-bound state'
configurations were found,
in which the pions remained trapped on timescales many times larger than
the light-crossing time of the initial configurations. In future work
we hope to understand their dynamics better, as well as the prospects 
for their detection through correlation measurements.

Computational limitations prevented us from performing complete 
investigations of the case of $k > 2$. While we can perform individual
runs, the time required to compute the associated probability 
distribution for $f$ was prohibitive. 
Nevertheless, for the simplified cylindrical configurations we considered
it is plausible that the $k = 2$ case captures the correct qualitative 
behavior for larger numbers of domains, as in any individual run the
evolution is seen to be dominated by nearest domain interactions.
Our results suggest that
even events with a large number of small domains are capable of 
providing a detectable signature. We are hopeful that analysis of
future RHIC data will allow  stringent upper bounds to be placed on the
probabilities of multiple domain formation, or alternatively the
discovery of coherent phenomena in heavy ion collisions.

\bigskip
\noindent

The authors would like to thank John Harris, Rudy Hwa, 
Vincent Moncrief, Krishna Rajagopal and Dirk Rischke 
for useful discussions and comments. Nick Evans and Stephen Selipsky
are acknowledged for early participation in this project.
This work was supported in part under DOE contracts DE-AC02-ERU3075 and
DE-FG06-85ER40224.

\bigskip

\section{Appendix A : Relations Between $\pi$ and $\phi$ Fields}

The $\phi$ and $\pi$ fields are related via 
\beq 
\phi _0 + i \vec\tau \cdot \vec\phi =
F_\pi e^{i \vec\pi \cdot \vec \tau / F_\pi} ~~.
\eeq
Using the identity
\beq
e^{i\vec A \cdot \vec \tau} = \cos(|A|)+i {\vec A \cdot \vec \tau \over |A|}
\sin(|A|)
\eeq
we can write out the above equation
\beq
\phi _0 + i \vec\tau \cdot \vec\phi =
F_\pi \cos({|\pi| \over F_\pi})+i F_\pi {\vec \pi \cdot \vec \tau \over |\pi|}
\sin({|\pi| \over F_\pi})
\eeq
which implies
\beq
\phi_0 = F_\pi \cos({|\pi| \over F_\pi}) \label{pi2phi0}
\eeq
and
\beq
\vec \phi = F_\pi {\vec \pi \over |\pi|}\sin({|\pi| \over F_\pi})~~. \label{pi2phi}
\eeq

By inverting the equation for $\phi_0$ above, we get
\beq
|\pi| = F_\pi \arccos({\phi_0 \over F_\pi})~~,
\eeq
and since $\vec \pi$ and $\vec \phi$ point in the same direction (as seen
from the equation solving for $\vec \phi$ above), we get
\beq
\vec \pi = F_\pi \arccos({\phi_0 \over F_\pi}) {\vec \phi \over |\phi|}~~.
\eeq
However, this will not suffice for the case where the values of the $\phi$
fields are approximated.  The problem is that there are four $\phi$ fields
while there are only three $\pi$ 's.  The reason that this is possible
is because there is an implicit condition on the $\phi$ 's -- that
$\phi^2 \equiv \phi_0^2 + \vec \phi^2 = 1$.  If this is broken even
slightly, it is possible that $\phi_0 > F_\pi$, and then
$\cos^{-1}({\phi_0 \over F_\pi})$ becomes imaginary.

To avoid this problem, a fourth parameter is coupled with the $\pi$'s,
a $K$ field that measures the length of $\phi$ ($K^2 = \phi^2$).
Using this, equations~\ref{pi2phi0} and~\ref{pi2phi} become
\begin{eqnarray}
\phi_0 &=& K \cos({|\pi| \over F_\pi}) \\
\vec \phi &=& K {\vec \pi \over |\pi|}\sin({|\pi| \over F_\pi})~~.
\end{eqnarray}
Inverting these yields
\beq
K = |\phi| ~~,
\eeq
\nopagebreak
and
\nopagebreak
\beq
\vec \pi = F_\pi \arccos({\phi_0 \over |\phi|})
{\vec \phi \over |\vec \phi|}~.
\eeq
Now we can assume that $K \simeq F_\pi$ and throw away the
information in $K$.

\section{Appendix B : The Particle Detector}

The number of particles can be detected by using the usual Klein-Gordon number operator
in the limit of the simulation where the wave is dispersed and the nonlinearities have 
become negligible.  The number operator is

\beq
N = {1 \over (2 \pi)^3} \int d^3 \vec p ~~ a_{\vec p}^{\dagger} ~ a_{\vec p}~~.
\eeq

As we are in the classical limit, we can treat all these operators as 
c-numbers and simply 
solve the above in terms of the fields.  To do this we start with 

\beq
\pi (x) = \int {d^3 \vec p \over (2\pi)^3}{1\over \sqrt {2E_{\vec p}}}
~(a_{\vec p} ~ e^{-i~p\cdot x}
+ a_{\vec p}^\dagger ~ e^{i~p\cdot x}) ~ .
\eeq

After some algebra, this equation can be inverted to solve for $a$, yielding

\beq
a_{\vec p} =
{1 \over \sqrt 2} e^{iE_{\vec p}}
\int d^3 \vec x
~\left(\sqrt{E_{\vec p}}~~\pi(x)
+ i\sqrt{1\over E_{\vec p}} 
~ \dot \pi(x)\right)
~e^{-i ~\vec p \cdot \vec x}
\eeq

\beq
= {1 \over \sqrt 2} e^{iE_{\vec p}}
~{\rm{FT}}_{\vec p}\left[ \sqrt{E_{\vec p}}~\pi(x)
+ i\sqrt{1\over E_{\vec p}}~\dot \pi(x) \right] ~.
\eeq

It might seem natural to simply insert this into the definition of N
and derive an explicit expression. However complexities 
arise due to the non-local nature of the N operator. Hence, we separately calculate the
$a_{\vec p}$ field then use this to obtain N.

Two useful equations are those of the Fourier transform in axial and spherical 
coordinates, given by the spherically symmetric case

\beq
{\rm{FT}} (P_r)[f(r)]=
\int d^3{\vec x}~f(r)e^{-i~\vec x \cdot \vec p}
= 4 \pi {1 \over p_r} 
\int _0 ^\infty dr ~f(r)~ r ~ \sin(r P_r)
\eeq
and the axially symmetric case
\beq
{\rm{FT}} (P_r, P_z)[f(r, z)]=\int d^3{\vec x}~f(r, z)~e^{-i~\vec x \cdot \vec p}
= 2 \pi \int _0^\infty \int _{-\infty}^\infty
dz ~dr ~f(r)~ r ~J_0(r P_r)~e^{-i~z P_z} ~ .
\eeq

\begin{figure}[htb]
\center{
\leavevmode
\epsfbox{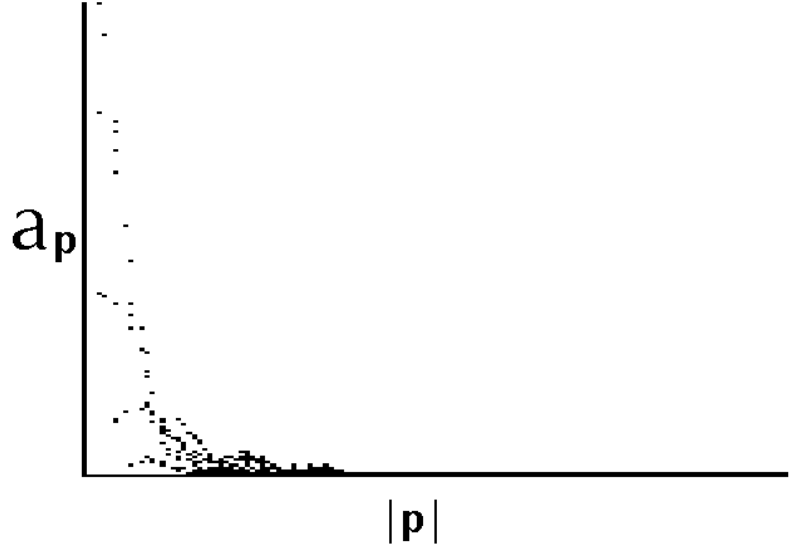}
\caption{An example of typical values for $a_p$ vs. $|p|$} \label{apvsps}
}
\end{figure}

In addition to its use in calculating the number of particles, transforming to momentum space
allows us to inspect against possible aliasing problems.  Because of the axial 
(or spherical) symmetry
and complicated shapes involved, this would be difficult to do analytically. However, a plot of
$a_p$ vs. $|p|$ (for representative values of $\vec p$) can be visually 
checked.  As in one dimensional 
Fourier Transformations, discretization copies the primary Fourier transformed image centered 
around zero momentum in a repeating pattern at higher frequencies 
(albeit deformed).  By viewing $a_p$ we insure that
each copy decays fast enough as to not interfere with its neighbor.  Figure~\ref{apvsps}
shows a sample $a_p$, which can visually be seen to decay before the 
second image (which is not shown).

\epsfysize=8.0 cm

\section{Appendix C : Rotating the Configuration}


To minimize the number of simulation runs, we used the 
results of one run to calculate the detected pion distributions
resulting from initial conditions which differ from those of the
first run by a rotation in isospin space.
The definition of the Klein-Gordon number operator is

\beq
N = {1 \over (2 \pi)^3} \int d^3 \vec p ~a^{\dagger}_{\vec p}~ a_{\vec p}
~~.
\eeq
In our case we have three associated number operators, for $\pi_{0,+,-}$:
\beq
N^j = {1 \over (2 \pi)^3} \int d^3 \vec p ~(a^j_{\vec p})^{\dagger} ~ a^j_{\vec p}~~.
\eeq

An isopsin rotation on the configuration yields a new
set of $a_j$'s via the usual transformation,
\beq
\tilde{a}^j = R_{j'}^j ~ a^{j'} ~~,
\eeq
where R is a $3\times 3$ $SO(3)$ matrix.
Applying this to the $N_i$'s, one obtains an expression
involving cross terms $(a^j)^\dagger a^k $, so a new tensor is introduced
\beq
N^{j, k} = {1 \over (2 \pi)^3} \int d^3 \vec p ~(a^j_{\vec p})^{\dagger} ~ a^k_{\vec p}
~~,
\eeq
where the single indexed $N^i$'s are just the diagonal elements of
the double indexed $N^{j, k}$,

Now the N's can be trivially rotated
via the usual tensor rotation law, 
\beq
\tilde{N}^{j, k} = R_{j'}^j ~ R_{k'}^k ~N^{j', k'}~~.
\eeq

\section{Appendix D : Computational Technique}
\label{sec-comptech}

Besides initial configurations, each numerical simulation had a series of 
internal parameters, whose values were independent of the physics of the problem but needed to
be determined in order to insure proper results were obtained. 
These parameters included the spatial discretization $\Delta x$, 
time step $\Delta t$, total evolution time $T$, and linear sigma potential scaling $\lambda$.

Of these, $\Delta x$ and $\Delta t$ proved to be quite simple to fix, as the final results were
insensitive to the chosen values, aside from cases where evolution became obviously unstable.  
Chosen  $(\Delta x, \Delta t)$'s within the stable region yielded final values which
agreed within reasonable errors, so for a given $\Delta x$, the coarsest stable value of 
$\Delta t$ was 
chosen, thus 
decreasing needed computational power.  It should be noted that the required
$\Delta t$ decreased as field strength and $\lambda$ were increased to high 
values, because when this is done, higher frequency oscillations  
become more dominant in the evolution.

\begin{figure}[htb]
\center{
\leavevmode
\epsfbox{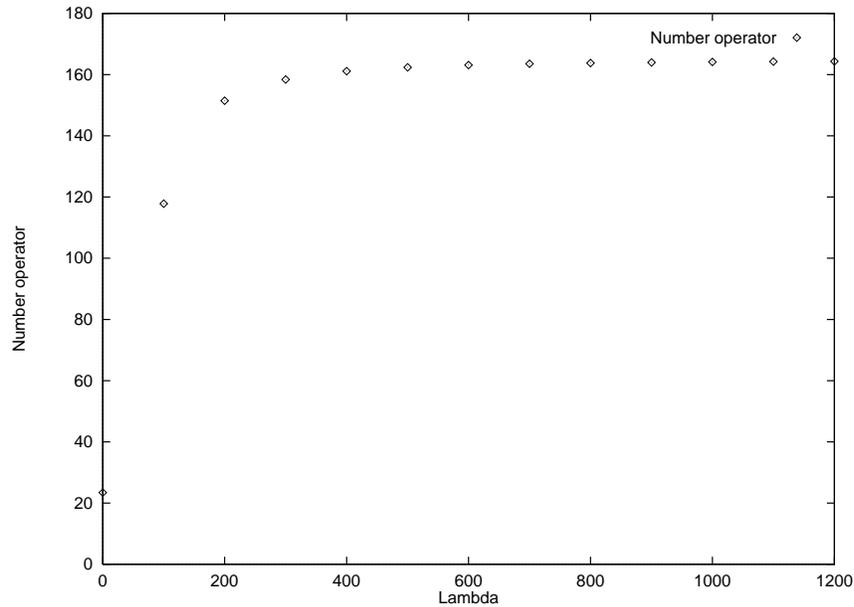}
\caption{Number operator vs. $\lambda$} \label{lambda}
}
\end{figure}

A reasonable value of 
$\lambda$ was found by holding everything else fixed, then
calculating and 
plotting the number operator for varying $\lambda$ (see figure~\ref{lambda}).  
From this plot we were able to visually 
pick out the convergence point of the curve.  
The curve converges for values of $\lambda$ in the hundreds, but as the computational 
cost of raising $\lambda$ turned out to be low, the much more conservative 
value of $\lambda 
= 6000$ was chosen.

\epsfysize=8.0 cm

\begin{figure}[htb]
\center{
\leavevmode
\epsfbox{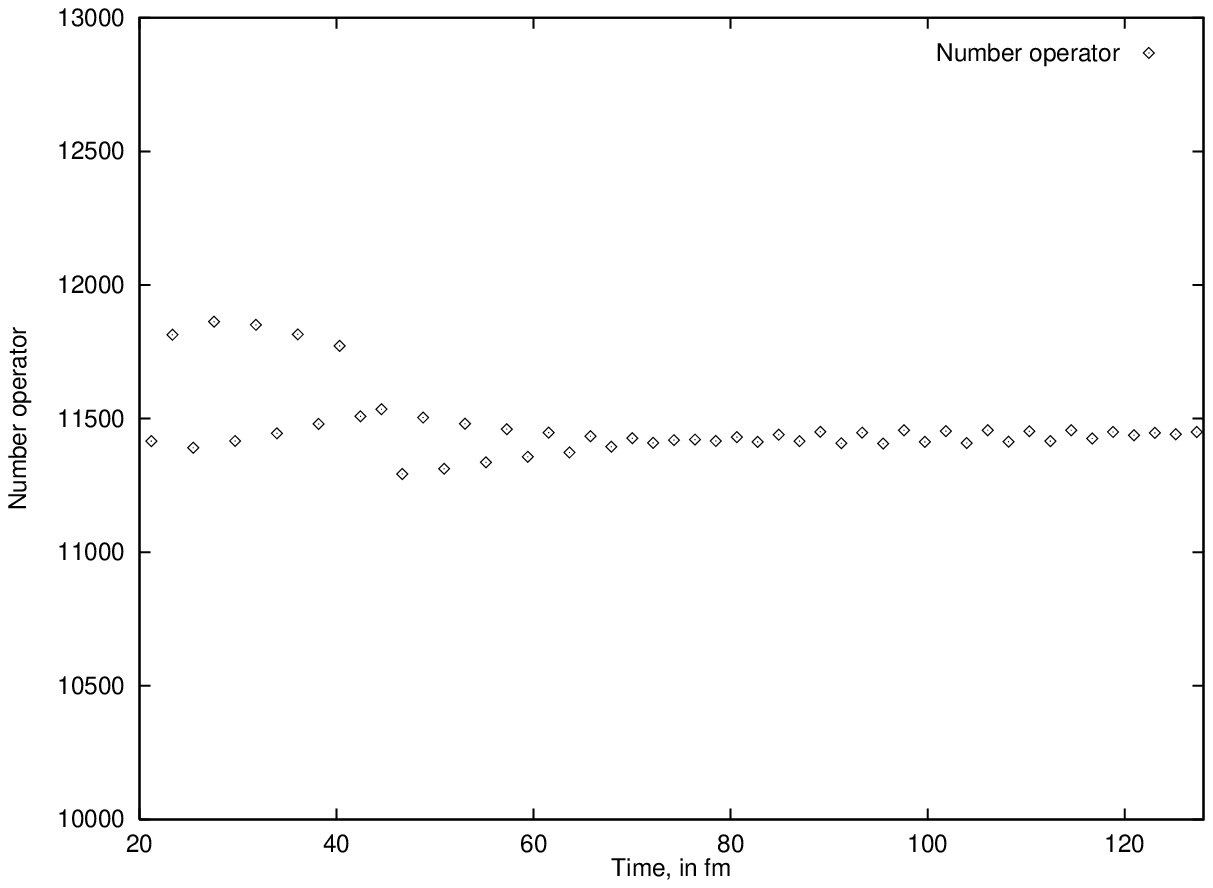}
\caption{Number operator vs. evolution time} \label{et}
}
\end{figure}

The remaining parameter, evolution time $T$, proved to be the least straghtforward to work with.  In 
principle the same 
technique was used as for $\lambda$, but in practice difficulties resulted due to the large 
and sudden variation in decay time 
for differing initial conditions (see 
section~\ref{nonlin} for details of this phenomenon).  Longer evolution times stretched 
the limits of our machines' memory and processor speed, and in the end were the deciding factor in 
limiting initial field strengths less than or equal to $1.6 ~ F_\pi$. 
Figure~\ref{et} shows a typical plot of calculated N versus $T$ illustrating how the convergence 
process took place.

\epsfysize=8.0 cm

\end{document}